\documentclass{article}

\usepackage{arxiv}

\usepackage[utf8]{inputenc} 
\usepackage[T1]{fontenc}    
\usepackage{hyperref}       
\usepackage{url}            
\usepackage{booktabs}       
\usepackage{amsfonts}       
\usepackage{nicefrac}       
\usepackage{microtype}      
\usepackage{lipsum}		
\usepackage{graphicx}
\usepackage{natbib}
\usepackage{doi}
\usepackage{fancyhdr}

\fancypagestyle{acm-foot}{\fancyhf{}\fancyfoot[L]{This paper has been accepted for publication in the Proceedings of the 2023 CHI Conference on Human Factors in Computing Systems, April 23--28, 2023 in Hamburg, Germany and has been assigned {DOI}: 10.1145/3544548.3580886}}

\begin{document}
\renewcommand{\thefootnote}{\fnsymbol{footnote}}

\title{Designing Ocean Vision AI: An Investigation of Community Needs for Imaging-based Ocean Conservation}


\author{
	\href{https://orcid.org/0000-0002-3610-1319}{\includegraphics[scale=0.06]{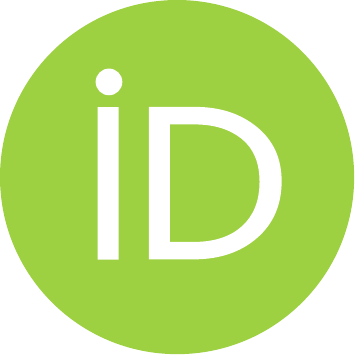}\hspace{1mm}Alison Crosby} \\
	University of California, Santa Cruz\\
	Santa Cruz, CA, USA 95064 \\
	\texttt{arcrosby@ucsc.edu} \\
	\and
    \href{https://orcid.org/0000-0002-9822-6774}{\includegraphics[scale=0.06]{orcid.pdf}\hspace{1mm}Eric C.~Orenstein} \\
	Monterey Bay Aquarium Research Institute\\
	Moss Landing, CA, USA 95039 \\
	\texttt{eorenstein@mbari.org} \\
 	\and
    \href{https://orcid.org/0000-0002-2514-7100}{\includegraphics[scale=0.06]{orcid.pdf}\hspace{1mm}Susan E.~Poulton} \\
	Ocean Discovery League\\
	Saunderstown, RI, USA 02874 \\
	\texttt{susan@oceandiscoveryleague.org} \\
 	\and
    \href{https://orcid.org/0000-0003-0826-0758}{\includegraphics[scale=0.06]{orcid.pdf}\hspace{1mm}Katherine L.C.~Bell} \\
	Ocean Discovery League\\
	Saunderstown, RI, USA 02874 \\
	\texttt{croff@alum.mit.edu} \\
 	\and
    \href{https://orcid.org/0000-0001-9039-3255}{\includegraphics[scale=0.06]{orcid.pdf}\hspace{1mm}Benjamin Woodward} \\
	CVisionAI\\
	Medford, MA, USA 02155 \\
	\texttt{benjamin.woodward@cvisionai.com} \\
 	\and
    \href{https://orcid.org/0000-0002-3434-9353}{\includegraphics[scale=0.06]{orcid.pdf}\hspace{1mm}Henry Ruhl} \\
	Central and Northern California Ocean Observing System\\
	Moss Landing, CA, USA 95039 \\
	\texttt{hruhl@mbari.org} \\
  	\and
    \href{https://orcid.org/0000-0002-7249-0147}{\includegraphics[scale=0.06]{orcid.pdf}\hspace{1mm}Kakani Katija} \\
	Monterey Bay Aquarium Research Institute\\
	Moss Landing, CA, USA 95039 \\
	\texttt{kakani@mbari.org} \\
  	\and
    \href{https://orcid.org/0000-0002-8700-7795}{\includegraphics[scale=0.06]{orcid.pdf}\hspace{1mm}Angus G. Forbes} \\
	Purdue University\\
	West Lafayette, IN, USA 47907 \\
	\texttt{agforbes@purdue.edu} \\
}

\renewcommand{\shorttitle}{Crosby et al.}

\hypersetup{
pdftitle={Designing Ocean Vision AI: An Investigation of Community Needs for Imaging-based Ocean Conservation},
pdfauthor={Crosby et al.},
pdfkeywords={HCI design and evaluation methods, Environmental sciences, Artificial intelligence},
}

\maketitle
\thispagestyle{acm-foot}
\begin{abstract}
Ocean scientists studying diverse organisms and phenomena increasingly rely on imaging devices for their research. These scientists have many tools to collect their data, but few resources for automated analysis. In this paper, we report on discussions with diverse stakeholders to identify community needs and develop a set of functional requirements for the ongoing development of ocean science-specific analysis tools. We conducted 36 in-depth interviews with individuals working in the Blue Economy space, revealing four central issues inhibiting the development of effective imaging analysis monitoring tools for marine science. We also identified twelve user archetypes that will engage with these services. Additionally, we held a workshop with 246 participants from 35 countries centered around FathomNet, a web-based open-source annotated image database for marine research. Findings from these discussions are being used to define the feature set and interface design of Ocean Vision AI, a suite of tools and services to advance observational capabilities of life in the ocean.
\end{abstract}

\keywords{ocean science, sustainability, data portals, machine learning-enabled interfaces, citizen science, human-centered design}

\maketitle

\section{Introduction}

Discovering what species exist in the ocean and their distribution across different regions is a daunting challenge~\cite{Haddock2017}. The ocean is filled with life that we have yet to describe and is governed by numerous chemical and physical processes that ocean scientists are only beginning to understand. Studying organisms in the ocean with traditional, resource-intensive sampling methodologies limits the ability of researchers to resolve vitally important biological-physical interactions and engage diverse communities~\cite{lehman2018ships}. However, with the use of modern robotics technology, low-cost observation platforms, and distributed sensing tools, ocean scientists are developing methods to find new animals and unravel the complex relationships that govern their lives~\cite{zhang2015future,wang2019advancing,piermattei2018cost}. Some scientific communities have made progress in scaling their observations using distributed platforms and open data structures. For example, the chemical and remote sensing communities gather data using satellite remote sensing of near-surface ocean conditions and via the Argo Program's global float array. However, large-scale analysis of marine biological communities and ecological processes has largely lagged behind~\cite{claustre2020observing,mckinna2015three,muller2018satellite}.

The ocean is a uniquely challenging environment in which to study and monitor the inhabitants that call it home. Scientists estimate that there are on the order of 1 million marine species in the ocean and that as much as 60\% of them are totally undescribed~\cite{appeltans2012magnitude}. Even the degree to which marine organisms are understudied is likely itself an underestimate~\cite{lin2022estimating}. Marine professionals are now turning to digital imaging systems to discover, study, and monitor the denizens of our seas. In situ image- and video-based sampling of biological communities enables the identification of animals to the species level, elucidates community structure and spatial relationships in a variety of habitats, and reveals the fine-scale behavior of organismal groups~\cite{durden2016perspectives,lombard2019globally,walcutt2020assessment,langlois2020field,Katija2020}. Underwater ecological surveys with imaging technology have become increasingly tenable due to the ease with which digital systems can be deployed and the availability of remotely controlled and autonomous platforms to carry them~\cite{giering2020sinking,greer2020high}. Imaging is also an effective engagement tool, giving broader communities access to marine life and insight into issues facing the ocean~\cite{markowitz2018immersive,fauville2020virtual}. Unfortunately, processing visual data, particularly data with complex scenes and containing organisms that require expert classifications, remains a resource-intensive process that is not scalable in its current form~\cite{van2015building}. 

Current estimates indicate that well over 300,000 hours of underwater video footage have been collected globally to date and that less than 15\% of it has been annotated by human experts~\cite{Bell_ODL_Capacity2022}. The rate at which such data is being collected is increasing every year, adding to an already extensive backlog. Moreover, these numbers do not include the enormous volume of still images and microscopy data that is regularly gathered. There is a clear need to develop effective automated strategies to assist human experts and enthusiasts in their efforts to use this invaluable repository. Unsupervised learning methods have been used to identify regions or moments of interest in underwater video footage during vehicle deployments and post-processing annotation tasks~\cite{zurowietz2018maia}. Supervised learning methods, trained on visual data where all objects have been identified (i.e., classified and localized), have proven effective for automating tasks to the genus and species level~\cite{lombard2019globally,chen2021new}. However, these machine-learning algorithms require access to large image-labeled training sets in order to achieve high accuracy across a diverse range of taxa. Given the potential of these algorithms, the underwater imaging community has called out the need for publicly available, comprehensive, large-scale image training sets, image and video analysis workflows, large-scale community-based verification, and rapid data analysis and export to data repositories and projects for subsequent scientific analysis. By creating such a pipeline, ocean scientists hope to enable accurate, accelerated processing of underwater visual data. Such a globally integrated network is critical for scientific inquiry, to inform all sectors of what is often referred to as the ``Blue Economy''--- ocean-related industries and resources that play a central role in climate mitigation strategies, renewable energy generation, and sustainable food harvesting and culturing--- and to ensure effective marine stewardship~\cite{bennett2019towards}.

In this paper, we investigate community needs around ocean visual data sharing, visualization, and human and automated annotation. The ocean science community has devoted a great deal of effort to developing annotation interfaces and training machine learning models~\cite{gomes2016current,zurowietz2018maia, pedersen2019detection, ellen2019improving}. But such work has largely been conducted by marine researchers seeking to solve a specific data problem for a narrow set of applications. There has, to our knowledge, never been a concerted effort to apply human-centered design principles to the marine imaging space to align stakeholder needs across a spectrum of use cases related to scientific exploration, biodiversity surveys for site management, and public engagement with marine organisms.

In the following sections, we describe our process of engaging with a diverse group of stakeholders representing the activities of the many sectors that comprise the Blue Economy, including fisheries, scientific laboratories, government agencies, and non-governmental organizations. Through a series of interviews and workshops with participants from around the world, we identified twelve user “archetypes’’ who we expect to engage with machine learning-enabled imaging tools in various capacities. Our analysis of the data collected through these interviews enabled us to better understand the functional requirements of these archetypal users engaged in ocean conservation and confirmed the importance of machine learning approaches for a range of ocean conservation tasks. At the same time, it also led us to identify issues related to accessibility, data sharing, community engagement, and the unequal distribution of both expertise and resources that currently inhibit the development of imaging-based ocean science analysis tools. In addition to enabling the creation of more powerful machine-learning enabled tools to support the activities of ocean conservation communities, we argue that the use of human-centered design methods is essential for developing effective and equitable approaches that advance scientific research and promote sustainable ocean-based economies that create jobs and support livelihoods across the globe~\cite{ODL_equity}.

\begin{figure}[h]
    \centering
    \includegraphics[width=1.0\linewidth]{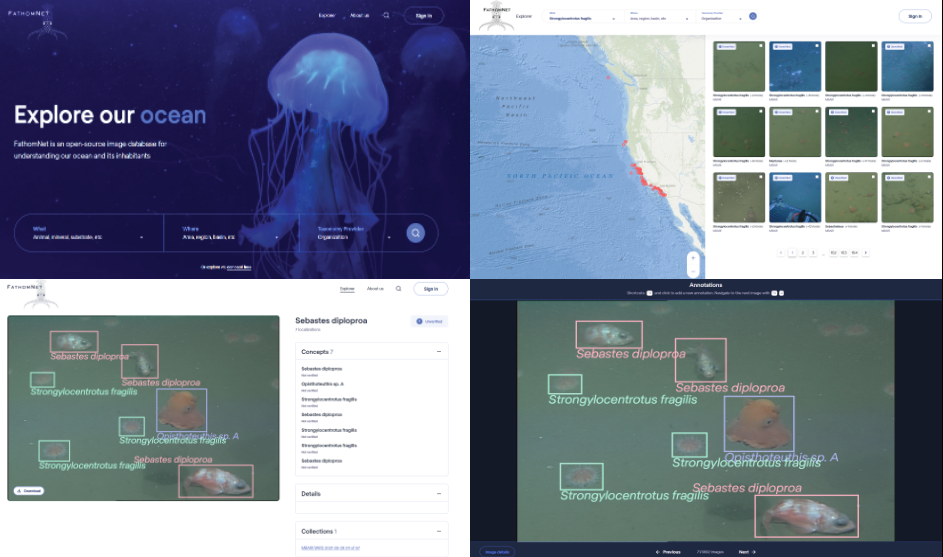}
    \caption{The FathomNet data portal contains features that include a simple search bar for terms in the concept tree, filtered searches where images can be displayed based on geographic location or terms within the concept tree, image display pages where concepts, details, and contributors' information is shown, and basic annotation and localization tool to allow users to augment or correct uploaded data in the database.}
    \label{fig:fathomnet}
\end{figure}

\begin{figure*}[h]
    \centering
    \includegraphics[trim={0cm 17cm 35cm 0cm},clip,width=1\textwidth]{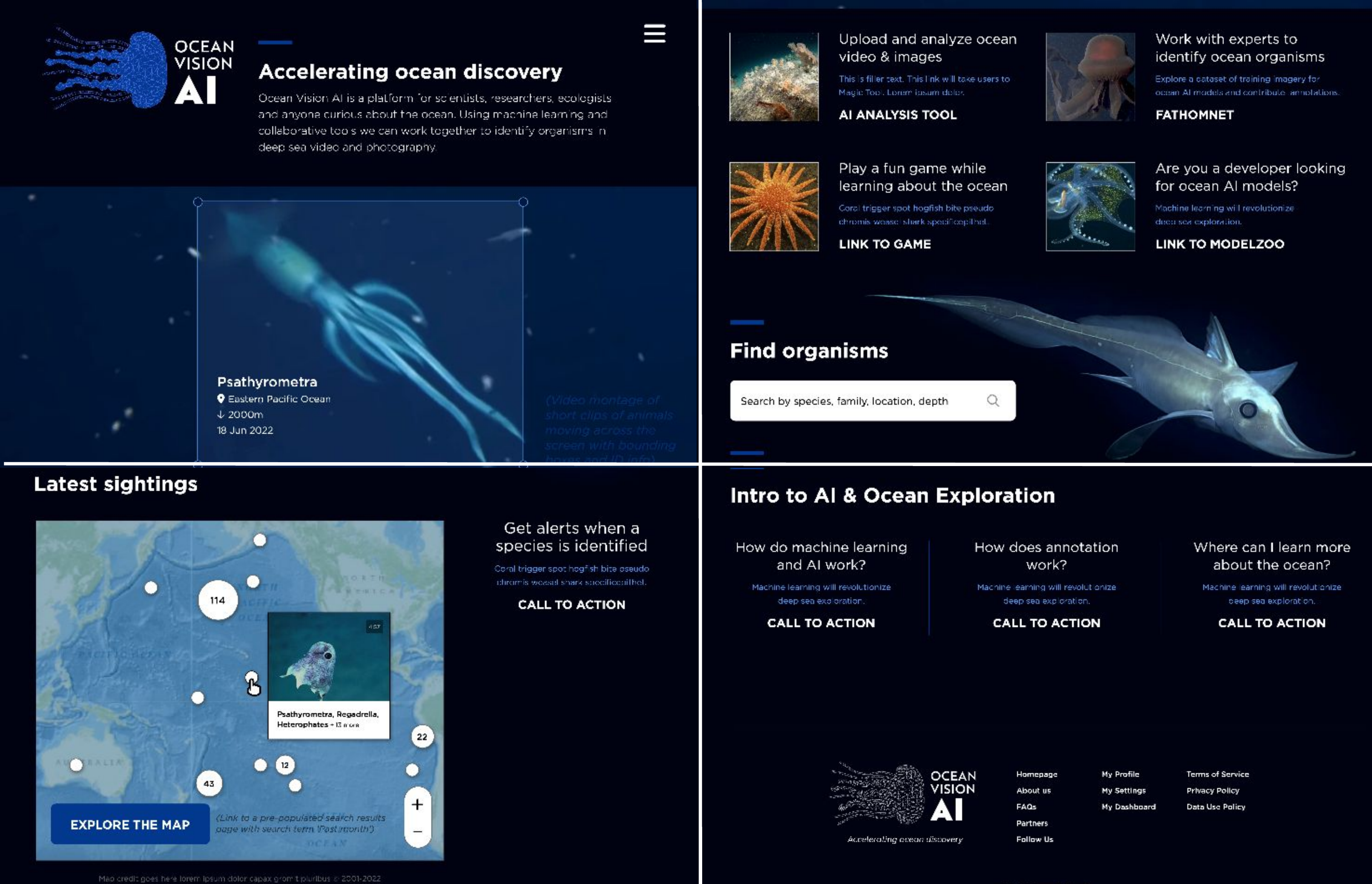}
    \caption{OVAI Portal home page. The vertical webpage is displayed here in separate panels from left to right, top to bottom. The OVAI Portal serves a wide range of users, including both experts and enthusiasts across multiple areas of ocean conservation (see Section 4.2), and provides functionality to make data analysis more accessible, simplify the training of classification models, encourage the uploading of new data, and to help users become familiar with taxonomic concepts. This functionality is integrated with the FathomNet database and is made available through games, data visualizations, tutorials, and search tools, among other modules.}
    \label{fig:portalhome}
\end{figure*}

The knowledge gained through understanding our user communities has directly informed the creation of a suite of machine learning-enabled tools, resources, and techniques--- collectively titled Ocean Vision AI, or OVAI--- for imaging, labeling, analyzing, and sharing ocean video and image data. These tools are centered around two interrelated efforts: The FathomNet database~\cite{katija2022fathomnet}, an open-source image database for understanding our ocean and its inhabitants, and the OVAI Portal, a suite of web-based tools leveraging FathomNet to conduct end-to-end analysis of visual data. FathomNet seeks to aggregate underwater image training data for all 202,063 accepted species in Animalia found in the WoRMS database using community-based taxonomic standards (Fig.~\ref{fig:fathomnet}). While we use species in the biological kingdom Animalia as an initial goal, the FathomNet concept tree can eventually be expanded beyond biota to include underwater instances of equipment, geographic and habitat features (including via existing ontologies), marine debris, as well as other taxonomic trees. The OVAI Portal enables the uploading and analysis of visual data by both researchers and enthusiast community members; the labeling and verification of annotations of species, including both identifications and localizations; and a range of functionality to support searching, querying, and exporting data (Fig.~\ref{fig:portalhome}). Responding to the needs of our interviewees and workshop participants, it provides users a straightforward interface to select concepts of interest, acquire relevant training data from FathomNet, and train machine learning models. Deployment of automated algorithms will result in ``annotation proposals’’ that can then be collaboratively verified by other community science contributors. Once data verification is complete, metadata (e.g., animal identifications, counts, timestamps) is provided to each project and exported to additional open data repositories that are used by specific communities. This direct connection with these recognized resources enables the broader ocean community to assess which animals are found where and when in the ocean, based on visual data alone. 

We intend this paper to make the following contributions: 1) We present a human-centered design approach to identifying and responding to the needs of diverse communities of ocean scientists, policymakers, and others engaged in ocean research and marine stewardship; 2) We synthesize a set of core findings from interviews and through feedback from workshop participants; 3) We identify a set of user archetypes participating in the Blue Economy and outline the ways in which they would make use of imaging-based data tools and data repositories. Broadly speaking, our contribution can be viewed as an instance of bridging the theory-practice gap as we develop a system that supports a spectrum of ocean stakeholders in their everyday practices~\cite{silberman2014sustainableHCI,remy2015theory}. While this project centers on the creation of analysis tools for ocean science data, we believe our work can serve as a model for ethnographic HCI studies that seek to inform design decisions for other ``big science’’ pipelines. Section 2 surveys and summarizes previous and current citizen science projects, machine learning techniques that support conservation and sustainability, and open-source imaging-based ocean science tools. Section 3 describes our human-centered approach toward identifying community needs for ocean conservation, including through our interview process and through feedback from our technical workshops. Section 4 provides our detailed findings regarding the user archetypes that need to be supported and describes the core issues that developers of imaging-based tools need to resolve in order to be successful. Section 5 presents a discussion of our findings and articulates the ways in which our investigation of community needs has informed the development and design of the Ocean Vision AI collection of tools. 

\section{Related work}

\subsection{Crowdsourcing tools for citizen science}

A number of existing tools have been created to enable the participation of citizen scientists in research endeavors, mainly in the area of data collection. For example, iNaturalist encourages users to snap photos of living creatures and upload them to a global database, which can then be accessed by trained scientists to support research projects that can benefit from these distributed observations~\cite{van2018inaturalist}. The iNaturalist team also developed the SEEK mobile app that uses an ML model trained on the iNaturalist data set to identify organisms in images captured on the device's camera. The user is encouraged to post images that the model cannot identify to iNaturalist itself~\cite{inat2020seek}. Similarly, Cornell Lab of Ornithology’s eBird project provides an interface to capture observations of different bird species, along with tools to assist in labeling the bird species based on the location of the observation~\cite{van2015building}. Zooniverse is another crowdsourcing platform that provides an infrastructure through which scientists can ask users to review and/or interpret data (mainly images) in order to assist in scientific analysis~\cite{simpson2014zooniverse,cox2015defining}. For example, WildCam Gorongosa is an ongoing project hosted on the Zooniverse that asks users to monitor footage from trail cameras and identify animals that they may find~\cite{nugent2017all}. Another Zooniverse-hosted project called Galaxy Zoo invites users to classify images of galaxies according to their shapes, helping astronomers to better understand how galaxies formed~\cite{masters2021galaxy}. In addition to answering questions to describe the shape of a galaxy, users can flag unusual features within the image, potentially enabling the user to contribute to new discoveries.

The tools provided by OVAI encourage enthusiasts to participate in the research process. Rather than asking users to collect data (as iNaturalist and eBird do), OVAI focuses on identification, labeling, and taxonomizing tasks, similar to the crowdsourcing activities available through Zooniverse. A key difference in OVAI is the interplay between the automated machine learning-powered labeling of images and the human interpretation of these images~\cite{torney2019comparison}. An important component of OVAI’s citizen science tools is that the user is encouraged to become educated in the process of labeling data according to a rich but rather complicated taxonomy of ocean species. That is, the citizen scientists are helping to train the machine learning networks while they themselves are becoming trained to become experts in understanding ocean data. Furthermore, the realities of scientific activities require that research-derived visual data may need to be controlled in a more nuanced manner than simply making it completely available (for example, datasets are often required by their organizations to be made accessible only to particular users, or are embargoed for a period of time before being made publicly available). 

\subsection{Image databases and machine learning tools for conservation research}

Tuia et al.~\cite{tuia2022perspectives} summarize challenges and opportunities in integrating machine learning algorithms in research pipelines, particularly those that leverage crowdsourced data platforms for conservation tasks. They note the mismatch between the drastic increase in data arising from advances in sensor technologies---remote sensors, camera traps, acoustic sensors, biologgers, and other monitoring devices---and our ability to analyze this data effectively. In particular, they note the possibilities enabled by machine learning: to detect and classify species; to identify (and re-identify) an individual member of a species; to detect or reconstruct the shape and pose of an individual in order to understand meaningful characteristics related to health and/or behavior; to reconstruct the environment where the species lives and model the diversity of that environment, including interactions within and between species. At the same time, they articulate ongoing challenges with integrating machine learning into conservation pipelines, including mitigating the inherent model biases that can arise in ecological datasets and the need for standards of quality control collaboratively established by model developers, researchers with domain expertise, and practitioners with local knowledge.   

There are many operational tools to assist image-based studies of terrestrial organisms both from a database and automated tools perspective. iNaturalist~\cite{van2018inaturalist} is one of the largest repositories of annotated imagery of animals and plants. It contains annotations that are crowdsourced from users, and there is a robust community of knowledgeable and skillful enthusiasts supporting users. Wildbook is an open-source software platform that helps researchers leverage automated tools to facilitate population analyses~\cite{berger2017wildbook}. It consists of a server-based, pre-trained object detection system and an interface to enable users to upload their images to Wildbook servers and interact with the output. The system cannot detect all organisms so the developers provide tools for manual annotations to create new models. For instance, Megadetector is a model that detects animals but does not label them, a workflow that demonstrably boosts human annotation speeds~\cite{norouzzadeh2021deep}.

These tools and efforts provide valuable insight into the development of marine-specific tools but do not necessarily provide an out-of-the-box solution. Many marine images are remarkably different in appearance from terrestrial ones, both due to organismal morphology and pixel level statistics of the images, stymieing direct application of many models~\cite{recht2019imagenet,hendrycks2021many}. Much of the ocean is also extremely difficult for many to access, limiting the number of pre-existing enthusiasts and potential citizen scientists to assist in annotation efforts.

\subsection{Ocean data management frameworks and annotation interfaces}

Despite the increasing number of large annotated ocean image datasets, to our knowledge, there are currently only two large-scale, extensible, publicly accessible data management frameworks. EcoTaxa is designed for segmented plankton images and is widely used by groups deploying the In Situ Icthyoplankton Imaging System and Underwater Vision Profiler instruments~\cite{picheral2010underwater,picheral2017ecotaxa}. CoralNet is a distributed coral database housing point-annotated images from around the world~\cite{beijbom2012automated,chen2021new}. These are excellent resources, but they are designed for a specific type of organism and thus have only been adopted by a subset of the ocean community due to their more narrow scope and engineering restraints related to image type. 

Many annotation interfaces have been developed for all manner of imagery, including visual data collected in the ocean~\cite{gomes2016current}. The ocean science community has built many data annotation interfaces, including BIIGLE, CATAMI, VIAME, and VARS, among others~\cite{ontrup2009biigle,harrison2012collaborative,dawkins2017open,schlining2006mbari}. These tools are open-source, typically supporting the use of contextual environmental metadata and often including some options for automated processing. Many of these tools are targeted toward a specific annotation project, and do not provide output annotations in a widely accepted format, and thus can be difficult for users who are not experienced programmers to set up~\cite{orenstein2022machine}. 

Several ocean-focused projects include public-facing community science or gamified components, including the Plankton Portal, NeMO-Net, and Deep Sea Spy~\cite{robinson2017tale,van2021nemo}. These tools were designed to present non-expert ``players’’ with imagery and guide them through annotation in an engaging way. The interfaces show players' unannotated raw data and ask them to sort it based on instructional videos, sometimes with the assistance of semi-automated leading questions. Though these types of interfaces have existed for several years, none have reached a large community of enthusiast users in the ocean space. 

\section{Methods}

Our methods were informed by a human-centered design (HCD) approach. HCD incorporates users at every step of the development process as designers aim to meet their needs~\cite{randall2007fieldwork}. Before developing a system, it is crucial for researchers to understand the entire context of a problem. Otherwise, design mismatches will occur and user needs and expectations will not be met. Designers rely on observational studies to minimize this mismatch and incorporate user participation throughout the development process. This approach enables a holistic view of a problem that incentivizes designers and researchers to see a problem from various perspectives. 

For citizen science projects in particular, Yadav and Darlington~\cite{yadav2016design} emphasize the importance of having
“a positive feedback loop between participation and learning and creativity'' 
in order to motivate scientists and volunteers. Yadav and Darlington follow a user-centered design approach to further analyze the participation of these two user groups, presenting guidelines for successful citizen science projects that are cost-effective, easy to maintain, trustworthy, and promote effective interaction between scientists and volunteers. Tinati et al.~\cite{tinati2015designing} note that for citizen science platforms to be successful, they need to both accomplish scientific objectives and to 
“attract and sustain the interest and support of a critical mass of volunteers over time''. 
They further note that, due to the scope of these two tasks, the process of creating effective tools can be unpredictable, even with the involvement of a competent team of design experts. Using Zooniverse as a case study for an analysis of citizen science platforms, Tinati et al. propose four focus areas that can help to mitigate this unpredictability, including  community development and task design. Much of their analysis was supported by and built upon previous design analysis work done by Kraut and Resnick~\cite{kraut2012building} on building online communities. These design considerations help to establish the benefits of engaging with and responding to the feedback of a platform's users. 

Related research that investigates designing crowdsourced platforms for the citizen science space further highlights the importance of incorporating users' perspectives. For example, researchers developing a paleontology app called FOSSIL~\cite{bex2018designing} first surveyed users' viewpoints to assess their needs and goals. This initial needs assessment informed the design and top priorities of the community. The authors additionally emphasize the importance of incorporating an iterative design process throughout the entirety of the application’s use to maintain engagement and alignment between the goals of users 
and scientists.

Notably, for ocean science, there have been few HCD efforts to inform larger user platforms. To our knowledge, only the recent work by Bell et al.~\cite{bell2022low} has explicitly incorporated an HCD approach for ocean technology design. They identified functional requirements for low-cost systems to facilitate deep-sea research worldwide via interviews with 20 marine professionals. Interviewees were also asked about their image and video analysis needs, particularly around an online machine-learning platform. The results illuminated several general design preferences for many ocean science researchers and managers, highlighting the ease of use, the ability to combine different data types, the utility of high accuracy at coarser taxonomic groupings, and the need for clear policies around data governance and ownership.

\subsection{Participants}

\begin{figure*}[h]
    \centering
    \includegraphics[width=.8\linewidth]{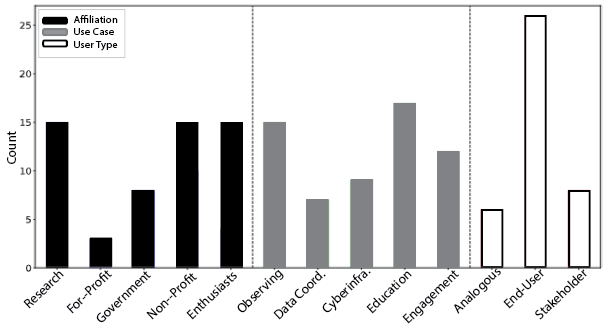}
    \caption{Interviewees were coded according to their affiliation, the type of use case they have for ocean visual data, and how they would use a system like Ocean Vision AI. Bar colors represent the different classifications. Interviewees often had relevant descriptions in each grouping (the sum of the bar heights in each group does not add to the number of participants).}
    \label{fig:intervieweebar}
\end{figure*}

We conducted 36 semi-structured interviews with professionals and enthusiasts within or analogous to the field of marine science over the course of four months. Participants were recruited through personal and professional connections of the project's collaborators, as well as references from the participants themselves. The interviewees were drawn from a diverse pool, hailing from 13 countries on every continent except Antarctica, of academic researchers, ocean enthusiasts, industry analysts, nonprofit advocates, government regulators, policymakers, and developers of analogous programs in different domains. We interviewed 25 individuals from the United States, 4 from Great Britain, 2 from South Africa, and 1 each from Australia, Germany, Japan, and Portugal. Together, they represent an array of potential OVAI users, contributors, and partners. We coded each interviewee's role in the Blue Economy, how they use ocean visual data, and how they relate to the OVAI program. An individual's role was assigned based on their professional or personal affiliations into five types: Academic or Research; Industry or For-Profit; Government; Nonprofit; or Enthusiast (Fig.~\ref{fig:intervieweebar}, black bars). Uses cases and interests were classified as: Observing; Data Coordination; Cyberinfrastructure; Education; and Engagement (Fig.~\ref{fig:intervieweebar}, gray bars). Finally participants were grouped according to how they relate to the OVAI program: \textit{analogous users} doing related work or research in a non-ocean science field; \textit{end-users} who would potentially use OVAI tools in their own work; and \textit{stakeholders} who may not interact with OVAI directly but have interest in seeing such an endeavor succeed and are supporting it specifically for their organization (Fig.~\ref{fig:intervieweebar}, white bars). Interviewees were often classified into several types within each grouping. For example, an interviewee could be affiliated with both research and government; interested in observing and data coordination; and interact with OVAI as both an end-user and stakeholder.  

\subsection{Procedure}

Interviews typically lasted one hour and took place over Zoom with two interviewers present, one taking on the role of facilitator and one acting as a notetaker. Before each interview, we asked for consent to record and informed them they could pause recording at any point. Zoom’s auto-transcription feature was also enabled for all interviews when applicable. The interviews were conducted in a semi-structured format with a prepared list of questions to ask each participant, but some of the questions varied slightly or were skipped entirely depending on the participant's professional background and experience. The template used for the interviews can be found below in the Appendix to this paper. 

In each interview, we introduced ourselves and asked the interviewee to tell us about themselves and their motivations for speaking with us. Next, we discussed key topics related to their interests and goals in ocean conservation and their use of technology in addressing those interests and meeting those goals. Example questions include: What topics related to your field do you care most about? What methods and techniques do you use to answer questions related to these topics? What hardware and software do you use? What are your needs for processing and generating underwater visual data? What observations would you like to achieve? How would you like to interact with and contribute to processing and generating underwater visual data? Does your group have any limitations in contributing to underwater visual data? Additionally, we solicited information regarding the interviewees' working community and their collaborator network, their communication practices, and how they were funded.

Finally, we provided details about the OVAI tools and asked a series of directed questions regarding the overall activities and goals of the OVAI initiative: How could a project such as this address your needs? Do you have any concerns about using artificial intelligence data in their applications (e.g., regarding regulatory requirements or intellectual property)? Would you or your organization have any concerns or hurdles using OVAI tools for their data (e.g., related to research embargoes)? We also asked the interviewee to envision their potential role in OVAI or a similar initiative.

\subsection{Data analysis}

The analysis used in our HCD approach derives from a use-inspired research curriculum developed in coordination with the design company IDEO~\cite{design2016field}. This design process is organized into three phases: \textit{Inspiration}, \textit{Ideation}, and \textit{Implementation}. The Inspiration phase involves framing your design challenge, creating a plan, building an interdisciplinary team, and then interviewing and observing the people you are designing for. The Ideation phase consists of reviewing the interviews, sharing memorable interviews with your team, isolating the top themes and ideas, synthesizing findings into statements, and exploring possible solutions through brainstorming, storyboarding, and/ or prototyping. The Implementation phase consists of piloting ideas with live prototyping, creating a timeline, assessing resources, and generally getting the project off the ground with funding and staffing. In our review of the interviews described in Section 4, we also followed a similar three-phase process, though we used a modified terminology (\textit{Information}, \textit{Illustration}, and \textit{Inspiration}) as our focus here was more on collecting information to both illustrate the need for our proposed tool and to inspire the tool design.

The data analysis used for our research consisted of a multi-step thematic analysis. At least two researchers first reviewed the transcripts and recordings of all interviews and defined each interviewer as either being an end-user, stakeholder, or analogous, so as to better identify needs based on the use cases of different types of participants. Members of our team then synthesized each interview by organizing key findings and quotes into groups related to the three phases of Information, Illustration, and Inspiration. This step allowed us to categorize discoveries and snippets into more accessible chunks of text, streamlining the review of the transcripts. For the Information section, we noted the background information of the participant and any other facts or details that helped us further our understanding (e.g., personal motivations). For the Illustration section, we noted any key anecdotes or quotes that either stood out or that helped to illustrate concepts related to OVAIs objectives. In particular, we also indicated whether there were any details or stories that depicted any current unmet needs of the participants. And lastly, for the Inspiration section, we noted new ideas or questions that could potentially be explored in future project phases. The Inspiration section was especially relevant for the analogous  participants and for participants from underrepresented areas of the ocean science community as their perspectives provide previously unrealized or overlooked concepts.  

After this initial synthesis phase, our team singled out key takeaways from each interview with post-it notes on a digital whiteboard using the Mural app. This occurred in a larger team setting where each interview was reviewed collaboratively by at least three researchers. The main takeaways (labeled on virtual post-it notes in Mural) of each participant were then grouped together based on the themes that emerged through this review process. Using these emergent themes (see Sec. \ref{sec:core-themes}) we noted the impact each had on the project and labeled the interview types that led to each theme. 

\subsection{Workshop}

\begin{figure}[h]
    \centering
    \includegraphics[trim={0cm 13cm 27cm 0cm},clip,width=1\linewidth]{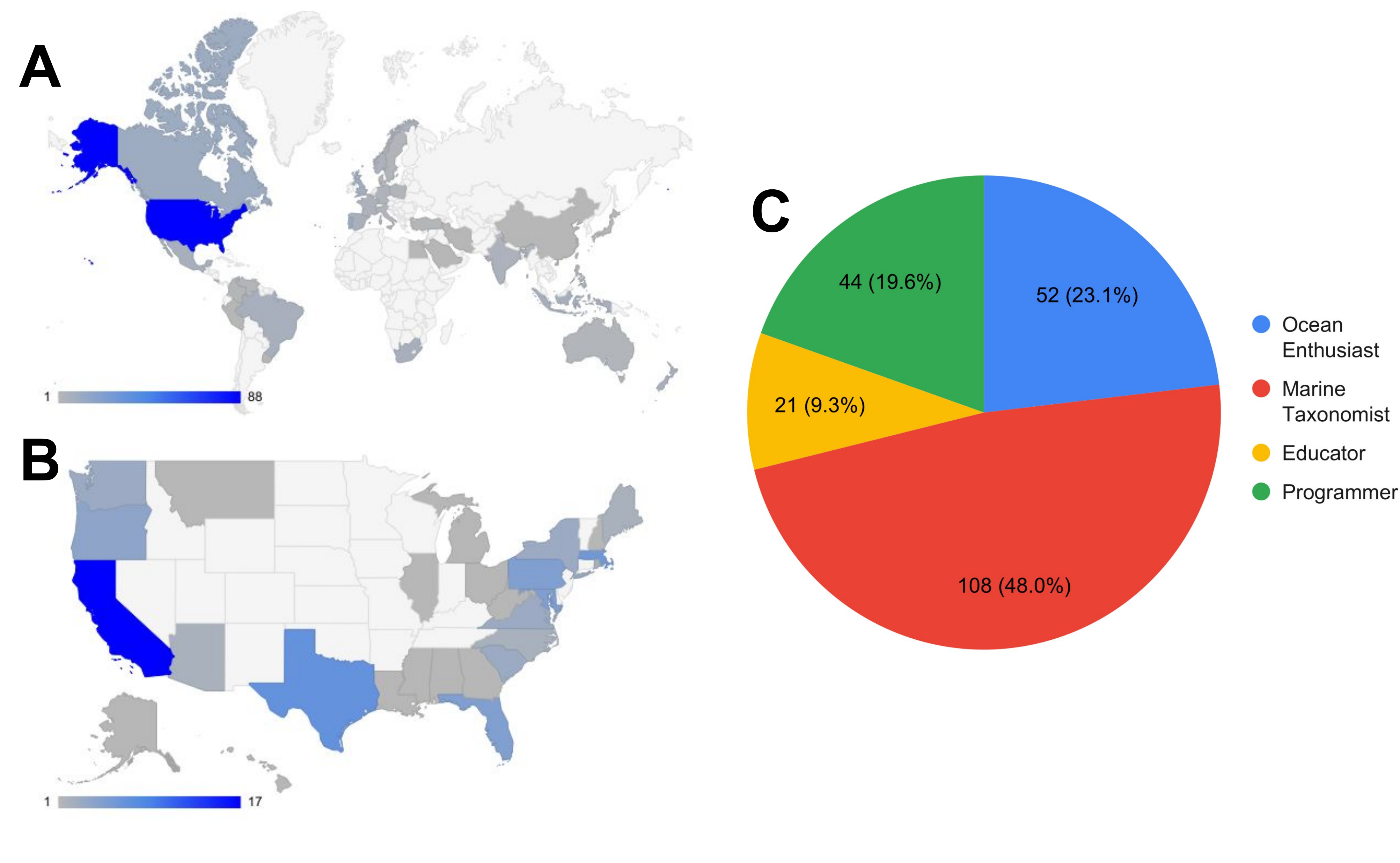}
    \caption{Attendees of the 2-day FathomNet workshop in April 2022 represented a variety of geographic locations and backgrounds. (A) Global and (B) national (mostly coastal) distributions of attendees. (C) Attendees came from a variety of backgrounds, including educators, programmers, ocean enthusiasts, and marine taxonomists.}
    \label{fig:fathomnetworkshop}
\end{figure}

In addition to interviews, our research team hosted a virtual workshop focused specifically on FathomNet, the web-based open-source image database for ocean research. FathomNet is a platform that enables users to train, test, and validate AI algorithms with curated datasets~\cite{katija2022fathomnet}. The 2-half-day workshop attracted 246 participants from 35 countries, representing stakeholders from every continent who are interested in using FathomNet for both their work and personal interest (Fig.~\ref{fig:fathomnetworkshop}). The first day of the workshop consisted of presentations and walkthroughs of different aspects of FathomNet. Participants learned how to filter annotations based on metadata, download existing annotated images, and upload new images and annotations. 

During the second day, attendees split into user-specific breakout groups: Educators, Programmers, Enthusiasts, and Marine Scientists. The group names were established by analysis of our interview findings, and each registered workshop participant was asked to self-assign themselves to one of the four groups. To ensure all voices and opinions were heard, the programmer, enthusiast, and marine scientists sections were broken into multiple sessions to limit the number of attendees to 10-20 per meeting. Across the two days, we had 21 participants who identified as Educators, 44 as Programmers, 52 as Enthusiasts, and 108 as Marine Scientists. Each breakout group participated in hands-on demonstrations of the existing FathomNet tools, via either a graphical or programmatic interface depending on interest. We then led brainstorming discussions for attendees to field suggestions regarding desired outcomes and needs from this type of platform.

The FathomNet workshop helped generate interest and name recognition among people interested in working with ocean visual data. These users were trained in how to use the tools and many volunteered to join quarterly working groups to stay involved. Participant feedback has informed ongoing development efforts of features targeting the different categories of users.

\section{Findings}

\subsection{Core themes} \label{sec:core-themes}

Through the iterative analysis process described above, we established 4 core themes: 1) There are several challenges with data sharing and community engagement in ocean science, 2) The ocean community has broad use cases for machine learning, 3) Overall there is very little machine learning and artificial intelligence knowledge within the ocean science community, and 4) There are no accessible tools for the ocean science community to process visual data using machine learning. Below we provide additional details and present excerpts from our conversations with participants that illustrate each of these themes. 

\subsubsection{Challenges with Data Sharing}

We discovered that a primary barrier to realizing OVAIs goals is the long-standing cultural norms against data sharing and community collaboration in the ocean space. This custom of keeping imagery ``in-house’’ limits progress. One particular aspect of keeping data in-house is that researchers fear that someone else will make a new discovery using their data before they can fully analyze it. This concern is more consequential for countries and organizations that lack the resources and people to complete the necessary analysis of their own data in a timely manner. For example, Participant 10 (end-user) reports on their fear of sharing data as a trust issue: 

\emph{``I think there is also a trust issue. As an emerging researcher, I need to publish, I want to understand my data [...] There’s also a trust issue with me, like okay I can get it there, but I also just really want to apply myself, my knowledge, my head and see what I get out of [the data]. And then everyone else can have it. Which is a mean thing to say, but that’s the reality. There’s a lot of trust dynamics within that. If I could store it and give it to a data manager that reassures me that no one is going to touch it, then maybe.”
}

Participant 21 (end-user) additionally points out an interoperability issue between their national marine research infrastructure and other organizations in their country as a data-sharing barrier: 

\emph{
``The overarching problem that we are having with all these different data sets it’s actually interoperability [...] Although we are the national marine information management system, we also have other departments and universities who collect data [...] Our system cannot communicate seamlessly with their system.”
}

Even beyond research groups, organizations that collect and share data publicly (as part of live streams) make it difficult for enthusiasts to locate both past and present video data. One member of an ocean science enthusiast group, Participant 5 (end-user), describes their difficulties: 

\emph{
``The footage is either unknown and also just not publicly accessible [...] NOAA for example has pretty good metadata and annotations associated with pretty much most of their dives, it’s just incredibly hard to find on their site though.”
} 

Any given set of visual data can support different projects, but few institutions have all the expertise and resources available needed to realize the full value of a dataset. In analyzing the interviews, we found that over half the interviewees consistently raised four problems: 

\begin{enumerate}
    \item No standardization of the imagery and associated metadata
    \item Inconsistent or inflexible taxonomy and concept trees
    \item No recognized metric for communicating inter- and intra-annotator quality
    \item Concern about attribution of data sources and recognition of taxonomic expertise
\end{enumerate}

\subsubsection{The ocean community has broad data use cases for machine learning}

There are numerous challenging use cases inherent in marine imagery that would appeal to machine learning research communities: biodiversity surveys require precise population estimates that must be robust to distribution shifts; species discovery needs automated systems that operate in an open world capacity; fine-grained taxonomic identifications could benefit from hierarchical model design. Participant 20 (end-user and stakeholder) shares an example of how ML could help with requests they get from researchers: 

\emph{
``We get requests from people studying octopuses, can you send me all of your video footage of octopuses? And it’s impossible to do because the data is all over the world. We get very specific requests for footage. So we want to try to break through that issue where people can just go on a website and download it for themselves.”} 

In this case, ML tools could make data more accessible to all and has great potential to increase the turnover of new research and discoveries. 

Another anticipated use case for ML is its potential ability to inform ROV (remotely operated vehicle) dive routes while scientists are at sea. Participant 22 (end-user and stakeholder) shares how they currently capture data at sea and how ML could improve their dives: 

\emph{
``If we don’t capture something in real-time, we’re probably going to miss it [...] If you’re doing that [annotating] in real-time, that can help inform what you’re doing in the moment, right? And so I think it could help inform an ROV dive that can last days potentially. That could help us make decisions on what we’re sampling, where we are going next.”
}

Additionally, 3 interviewees (2 end-user and 1 analogous) commented on the benefits of using machine learning for things like monitoring construction sites of offshore wind farms, surveillance data for fisheries, and ambient ocean knowledge for ocean farmers. Another interviewee (end-user) commented that AI is needed for observing and imaging the deep sea to help establish quantitative metrics to assess the vulnerability of certain areas. This use case is especially necessary to inform policymakers. 

The associated machine-learning solutions to these problems are all active areas of research that could benefit from access to data provided by sea-going imaging systems. The 6 computer scientists we spoke with recognize that utilizing this data would be beneficial but noted that it has not been made accessible to their community. These issues related to access lead to researchers feeling like it is more efficient to annotate their data by hand. 

\medskip

\subsubsection{Little machine learning/artificial intelligence knowledge within the ocean science community}

Ocean scientists, despite their wealth of visual data that might appeal to ML researchers, expressed frustration with the lack of available computational expertise. The first-order issue at the interface between these communities is a simple matter of money; ocean scientists and government managers typically cannot pay ML experts and engineers competitive salaries relative to those offered by the industry. This is especially true for marine scientists and policymakers from non-OECD countries. Participant 17 (end-user and stakeholder) highlighted the issue of competitive tech salaries causing institutions to train ocean scientists to be data scientists and programmers even though they are often not the best equipped and knowledgeable: 

\emph{
``We either don't hire the right people, the data scientists, or we aren’t training the people we do have. There are a lot of people like me that are fishery biologists that are then taking courses on how to develop an algorithm. And it’s like, well, that’s difficult. You’re probably not setting yourself up for the most successful program if you’re training fish biologists to run programs. We have that issue even with just data architects, database managers, software developers [...] We struggle to hire those modern types of positions into an agency full of fishery biologists.” 
} 

Participant 2 (stakeholder) discussed this lack of cross-domain within the ocean science field with the additive problem of having an AI position open for most of the past year with little hope of filling it: 
\emph{
``If you think about costs, what do you need to be able to develop an AI algorithm and apply it? You need Python, Python is free. You need a computer, most of these scientists already have a computer. It’s not about these physical costs, it’s your understanding of how to approach it [...] How to develop a training data set, how to train your algorithm, how to see whether your training has been sufficient or not. As soon as we start talking about these things, we’re losing 80\% of our scientists. These are all new concepts to them. It’s really hard for them to comprehend. [...] We've had an AI position open for the last 8 months and have not been able to attract anyone so far. This shortage of understanding AI principles, this shortage of AI expertise is what’s hurting us most.” }  

Without the financial ability to directly contract ML know-how, ocean scientists are left to either develop the expertise in-house or attempt to entice potential collaborators from academia or industry. Both paths are challenging especially when many ocean scientists are under the misconception that ML is a ``solved problem,” often leaning on students or trainees to implement workflows with little technical guidance, leading to poor results and disillusionment. 

\subsubsection{No accessible tools for the ocean community to process visual data}

Ocean-aligned researchers have a strong desire to integrate ML into their visual data pipelines as noted in key takeaways from 15 interviews. Indeed, all interviewees recognized that there is little hope of effectively using their data without ML assistance. Yet the prevailing perception is that there is no cohesive approach to do so in the ocean space. Many groups have portions of the workflow in place but have self-identified big gaps or deficiencies that seem insurmountable. Participant 1 (end-user and stakeholder) detailed aspects such as the lack of consistent image datasets and code bases to train systems on as being a major issue within the community:

\emph{ ``For every new use case, for every new dataset, we always have to tune systems again. So the system for nodule detection that I trained on this one dataset acquired with this one specific camera will probably not work right out of the box with the next nodule set acquired with a different camera. So changing the light, changing the camera, changing the deployment type, these all affect the image signal and this will affect the quality of any trained machine learning or just image analysis system. So if we can create something like ImageNet for the ocean where we can train systems that are agnostic of all these effects, that would be amazing. [...] We also don't really have a code base for everything. So we do a lot of stuff in C++ using OpenCV, we built some tools around that for some publications. But some PhDs prefer Python, of course there’s also OpenCV Python [...] We as scientists usually aim for the next paper and then afterwards we don't really care about that thing anymore. That’s also a reason why we don’t have a super large code base that we can share with the community. Everything is essentially separated and not integrated into one bigger system.”
}

Participant 12 also noted that the lack of standardized data practices requires additional work from researchers when trying to share their data: 

\emph{
``Data isn’t standardized. It’s not following data standards like Darwin Core and so there’s an extra step to get it standardized so that it can be shared and you know, I think that's really the limiting factor.” 
}

Three of the interviewees also identified the lack of data hosting solutions as a large hurdle they face. Most organizations, including government and industry players, do not have effective ways of making their imagery easy to share or work with internally. For Participant 16 (end-user), they described the frustration of working as a freelancer and not having access to a server to store all their data: 

\emph{
``One thing I find quite frustrating [with existing annotation services] is that you need to have a server for all of the data to be put on. And so for instance, because I spent the last 2 years as a freelancer, it’s like well, I don't have, for instance, an academic server.” 
} 

People and groups that generate data use all manner of storage, from external hard drives to personal computers to institutional servers, that are not public-facing. This makes it difficult to implement collaborative annotation frameworks, entrain community scientists, and entice computer vision researchers. 
The lack of consistent tools and data practices particularly hurts the accessibility of ocean science research at under-funded and less-established organizations around the world. Tools need to meet users at their level. Participant 10 (end-user) shares their challenges of having to perform in several different roles as being the root problem when discussing processing visual data:

\emph{
``Me as a researcher, as a taxonomist in South Africa, my love, my absolute joy is sitting in a basement looking at specimens. But I do not have that sort of privilege because I have to be the taxonomist, the AI reporter, policy developer, technician at sea, chief scientist. I have to be so many things […] It's important to understand those challenges when you start thinking: Where are we in terms of visual imagery? Why are we at this point when technology has developed over x-amount of years? Why are other places, other countries lagging along? And until we have that conversation it becomes quite difficult and very insensitive, to some degree, to talk about an output. Who is this output for?” 
}  

\subsection{User archetypes}

Through the interviews, we were able to collect opinions and better understand the needs of users across different communities who will interact with ocean data using OVAI tools. As such, the four themes described above additionally helped inform our process of creating 12 user archetypes of potential OVAI users. Below we describe our user archetypes and how each may use this system. We believe that this process is particularly important as it allows the creators of OVAI to address issues and support a variety of functionality that users require. Moreover, by understanding these needs, we can incorporate tools to enable members of different communities to better collaborate with each other.

\subsubsection{Ocean enthusiast} Passionate about marine organisms and eager to engage in a community of like-minded ocean lovers while learning about new animals. Wants to use their acquired knowledge to help scientists work with visual data either through direct annotation or via a video game interface. Ideally, there will be a mechanism for interaction with recognized experts both to validate individual annotations and help the enthusiast improve their work while making them more confident. 

\subsubsection{Academic biologist/ecologist} Has a range of desires and use cases for their collection of images and videos. This could include creating learning experiences for students, building capacity in parataxonomy, or using ML/AI for biodiversity surveys, ecological studies, or tracking species expansion. This user might not have all the vocabulary or computational resources necessary to make it happen. They want access to ML expertise, help with data storage and sharing, and easy-to-use interfaces for working with annotations and algorithms. The user or their organization might already have portions of an annotation workflow already in place.

\subsubsection{Academic computer scientist} Has no expertise nor particular affinity for ocean science though is happy to find ways to contribute to environmental causes. They want to access datasets to perform bleeding-edge ML experiments and publish the results in high-impact journals or conference proceedings. The datasets need to be packaged for easy access with explicit benchmarks for comparing algorithm performance. This user might respond to “challenges” or leaderboards. They might also be intrigued by video data for robotics experiments.

\subsubsection{Professional taxonomist} Wants an easier way to filter visual data to find rare or undescribed species. They largely want to remove “whitespace” with no interesting organisms so they can focus on creatures of interest. Ultimately, this user does not care about the particulars of the system nor whether or not it is automated; they just want something that works and has a flexible way of altering or tracking taxonomic designations. Public outreach and interaction with enthusiasts is a perk but not a prerequisite.  

\subsubsection{Non-profit scientist} Often has interests that align with academic researchers but typically with an eye toward informing and influencing decision-makers. They want to produce visualizations based on their biodiversity or ecological studies that can illustrate the effect of a policy change or the impact of a project. Sharing their visual data is both an effective way of using it for their surveys and engaging the public, perhaps garnering more support for their cause. 

\subsubsection{For-profit scientist} Often has interests that align with academic researchers but ultimately needs to use the data to inform company decisions or satisfy regulatory requirements. This user might need to adhere to company-imposed embargoes on sharing data but would perhaps be willing to pay more to use an effective service. Like their counterparts in the non-profit sector, they will place a premium on effective visualization tools. 

\subsubsection{Government scientist} Wants to leverage their visual data for stock assessments or management surveys. This user needs to be able to communicate their results to political appointees and the public. Their mission has lots of potential scientific output but is often more oriented to developing actionable metrics for policymakers. The use cases typically dovetail with other research communities.  

\subsubsection{Government program} Has a vast backlog of videos and images that they want to make easier for the public and other researchers to access. This user wants a central location so interested parties can access the government's data without having to hunt for it. They value interoperability and ease of use to facilitate the work of others. The hope is that workflows developed with their data will pay dividends in terms of increasing the analysis speed of future data collections. Certain government programs are also interested in developing the capacity to help with live annotation during data collection campaigns. 

\subsubsection{Non-profit organization} Has lots of visual data and nowhere to put it. This user might be attempting to make their data more public or to store it in a comprehensive but embargoed fashion. In either case, they need consistent data formatting requirements and one central location for their imagery to live. Like some government players, these organizations might want to implement live annotation interfaces for public use while their assets are in the water. 

\subsubsection{For-profit organization} Has lots of visual data that they want to store in a secure location that only they and their affiliates can access. This user wants employees to be able to easily share and collaborate on the same datasets. They will also want to rapidly produce and distribute reports among their ranks. Eventually, they might need to distribute the data to regulators or other stakeholders. 

\subsubsection{Journalist} Wants to find striking videos and images to accompany engaging stories. They primarily need an intuitive front-end interface to browse available data and perhaps be able to make inquiries of registered users, especially domain experts. 

\subsubsection{Media organization} Wants to find content for use in production. This user will need to easily communicate with individuals and organizations that generated the original visual data to secure consent for reuse. They may also want to connect with researchers with expertise in specific animals. 

\subsection{Workshop findings}

The FathomNet workshop helped identify themes specific to each distinct user group. The discussions and feedback have been used to further understand specific use cases and user needs. The key findings from the workshop are separated into 4 individual sections: 1) Educators, 2) Programmers, 3) Enthusiasts, and 4) Marine Scientists.

\subsubsection{Educators}

Educators were the smallest group in the workshop, with 21 people identifying themselves as ocean science educators. Attendees in this section were interested in connecting students with educational materials and meaningfully contributing to FathomNet. The educators identified specific learning goals for their classes--- such as exposing students to new concepts and providing an introduction to the deep sea--- that could be facilitated by using FathomNet’s real-world data in novel pedagogical ways. The educators hoped that using FathomNet data could yield educational outcomes like: increased interest in ocean biodiversity; greater awareness of marine science-related career paths; building enthusiasm and continuity within the community; and more general awareness of the marine environment and the many issues it faces. However, there are a number of limitations when it comes to onboarding new students and classrooms. For example, there is a lack of coordination and organized resources to assist with integrating something like FathomNet into curricula, and, in general, teachers are often stretched too thin to take on the additional effort this may require. 

Educators desire ``plug-and-play'' resources, such as websites that contain teaching modules and clear instructions for using them. Additionally, there is a need for clearly identifying helpful starting points and presenting how-to guides that reduce some of the barriers of entry for teachers. To enhance uptake, module writers should include comprehensive evaluation criteria. To integrate real data annotation into lesson plans, there must be an effective way to assess the quality of the annotations the students are making. This procedure would need to happen before uploading new annotations to the central repository for educators to feel comfortable using such a resource. 

\subsubsection{Programmers}
Programmers were the second smallest section with 44 participants, made up mainly of computationally-inclined ocean scientists interested in finding new ways to train and deploy models on their own data. Some attendees were software developers seeking to build tools on top of the existing Application Programming Interface (API) to service their organization's need for hosting, sharing, and annotating ocean image data. General interests and discussions focused on quality control for both human and machine annotations. Additionally, there was a desire for enabling different annotation techniques (segmentation, points) and imaging types (microscopic, stereo) within FathomNet. 

Programmers were interested in building a forum around FathomNet to share automated models trained for different organisms, camera types, and environments. ML model databases like Hugging Face make model sharing easy, but typically do not have easy-to-search metadata relevant to the ocean. A FathomNet ``model zoo''--- a community-maintained collection of supervised machine learning models trained on ocean image data--- could contain such codified metadata that would enable practitioners in the marine space to effectively bridge the gap to ocean scientists. More specifically, users suggested codifying existing metadata and adding new fields to specify performance on an internal validation dataset in an ecologically consistent manner. This extra information would improve a user's ability to differentiate between models and improve reproducibility. Formal requirements on file formats and descriptions will improve the interoperability of uploaded models in the long term. 

\subsubsection{Enthusiasts}
Enthusiasts made up the second largest section of the workshop with 52 attendants. The enthusiast section was a very diverse group of non-professional ocean scientists. They are individuals who do not work in ocean science-related fields but nonetheless have developed deep expertise specifically in marine biology and ocean taxonomies. Enthusiast groups have largely organically developed around live streams of ROV dives from ocean research vessels, such as the Schmidt Ocean Institute's R/V Falkor and Ocean Exploration Trust’s Nautilus Live. The interactions between enthusiasts generally take place in the YouTube comment section during a live stream, or on Twitter. Outside of live streams, discussions and interactions between enthusiasts occur on Discord servers and Facebook groups. Discussion revolves around taxonomic identification between members, though this identification process has not directly contributed to any scientific efforts. Recently, efforts have been made to use annotation platforms that allow enthusiasts to collaborate on annotations for both historic and real-time ocean video data. Overall, enthusiasts desired a way to meaningfully contribute to the visual data processing pipeline to more effectively support ocean science and discovery. They were also interested in connecting with other ocean enthusiast groups to leverage each of their strengths and skills. 

Community engagement was a critical theme discussed during the breakout sessions. Because enthusiasts would be contributing to annotations during their own time, it is important to understand their needs, desires, and limitations. Suggestions included: creating a reward system with points or other forms of recognition; allowing for anonymity for users who prefer not to be identified; and establishing a mentorship program and/or workshops to connect enthusiasts with domain experts. 

Some discussions focused on gamifying the annotation experience, for example, by allocating points based on the number and quality of contributions. This would enable users to both quantify their contribution and improve their expertise via direct feedback on their annotations. Points could lead to badges that show what level of expertise a user is in specific animal groups (e.g., level 6 in cephalopods but level 2 in geology). Other participants placed an emphasis on anonymity as they are often not comfortable with their level of confidence in identifying species and annotations. To better support enthusiasts, there is potential to establish a formal mentorship program. For example, an expert could mentor several enthusiasts contributing to their data analysis. Quick training workshops can also be set at specific intervals throughout the year to aid in the growth of enthusiast users. Enthusiasts have the ability to greatly help the ocean science community in getting through their backlog of data, as such, it is important to build a system that welcomes these users and that acknowledges their importance. 

\subsubsection{Marine Scientists}
The marine scientists' section was the largest with 108 attendees, which largely consisted of biologists, ecologists, and taxonomists who were interested in contributing their expertise to FathomNet. Scientists discussed improvements and additions to the platform, such as: how to verify a user's taxonomic expertise, how to manage and track disagreements in species identification, how contributors would receive credit for their identifications, and types of collaboration tools for users. They envisioned a platform like FathomNet as having many use cases: training students in marine organisms, collaborating across borders and around animals of interest, finding images for outreach and education purposes, and processing video data faster by using AI as a first pass. The themes within these session discussions included: communication between users, resources to ease contributions, ways to enable attribution, and quality control of annotations. 

Alleviating any hurdles to working on FathomNet is important to ensure the usability and longevity of the platform itself. The scientists expressed a desire for a searchable mapping between common names to the organism’s Latin name. Additional discussion points focused on ways of easing the uploading of data and simplifying the annotation process, such as by providing a template for CSV uploads to FathomNet, enabling storage of other localization types within metadata for annotations, allowing two-way contributions between FathomNet and other annotation tools (e.g., Tator, VIAME, BIIGLE, Squidle), and enabling high-level detections/localizations in an automated fashion. 

Ensuring users are credited for their work is a critically important feature to develop. Taxonomists in particular felt their expertise was undervalued and that they were often asked to undertake such labeling tasks with minimal incentive. FathomNet could facilitate this in several ways, primarily by generating digital object identifiers (DOI) for submitted collections of annotations. This would allow future users, scientists, and programmers alike, the ability to properly attribute the annotations to the correct expert. Contributions could also potentially be tracked at all stages (submitter, annotator, localizer, verifier, etc) if linked to a user’s account and the collection DOI. ORCID numbers could serve as a useful springboard for this sort of functionality since many marine scientists already use the system. Attendees also mentioned that their home institutions would look favorably on a “certificate of recognition” that would indicate participation in the FathomNet.

Overall, for FathomNet to be used reliably, there needs to be stringent quality control of the annotations. Attendees suggested that building a field for annotators to express their confidence level in their labels would be useful for sorting purposes. Enabling multi-tiered verification levels or confidence tiers would yield similar results. For example, if there is a sample that exists for that particular specimen, indicate whether the specimen was examined, has a sequence, or any open nomenclature tags. Another important aspect is verifying each user’s level of taxonomic expertise. Users could submit relevant biosketches that indicate research focus and level of expertise, length of time in the field, and their ORCID number. Verified experts could then nominate other users to be experts for different organisms or serve as mentors to less experienced users.

Largely, marine scientist users are hopeful that FathomNet could be a platform for a global marine taxonomy network. The community features the attendees suggested could be leveraged to create regional hubs of expertise, provide platform-specific training, and distribute resources for taxonomy workshops and events. There is additional interest in collaborative efforts that could be made by integrating FathomNet with other labeled datasets, like CoralNet and iNaturalist, as well as integrating pre-existing annotation tools.  

\section{Discussion}

Our interviewees and workshop participants come from different places, educational backgrounds, and professions. While they represent diverse ocean visual data use cases, their responses largely highlighted the parallel challenges they face. The themes we identified from our conversations and workshop survey are broadly consistent across the user groups. Opposing viewpoints and end goals between groups appear to be a function of nomenclature and terminology; discrepancies are more a matter of language rather than fundamental differences. The themes--- lack of data sharing, a desire for ML expertise in a domain science, a need for accessible digital tools, and data hosting--- are not unique to the ocean space. While these are common concerns found in data-intensive scientific research, they are worth bearing in mind and contextualizing for any large, public-facing initiative seeking to maximize utility for a specific domain challenge. To that end, our analysis of the user interviews and feedback from the workshop participants have led us to prioritize a set of functional requirements for the suite of Ocean Vision AI tools. 

We discovered that data sharing among individuals and institutions collecting in situ data is limited due in large part to a lack of standardization, inconsistent taxonomic concepts, and concerns about data quality and provenance. Meaningful standards must therefore be clearly implemented and communicated in every part of the system to enable consistent access and ensure data longevity. Taxonomic designations will need to be easily adaptable and will require a specific API. OVAI will need to deploy a coherent method for evaluating label quality and measuring variability among human annotators. Given OVAI’s community-oriented mission, a mechanism for awarding contributions must be in place to build a broad user base that entrains everyone from leading experts to lay enthusiasts. This could entail creating Digital Object Identifiers for data and identification efforts for academics or building membership lists and web pages to highlight achievements. These formal and informal methods would allow academics to track and cite their contributions while enabling enthusiasts to share and engage with ocean media. Allaying concerns surrounding the attribution of images, labels, and expertise will help curate a contributor network while also creating a collaborative space for users to work toward a common goal. 

We found that machine learning solutions were widely believed to be necessary for analyzing ocean visual data, but that these solutions were often not readily available to ocean science communities. Even when an individual or organization had access to resources, they were sometimes hard to access and difficult to use. Moreover, there was a perceived lack of computational expertise within ocean science communities. The costs of contracting a data science engineer are outside the budget for many organizations. Even when a marine expert learns how to run ML resources, they do not have time to familiarize themselves with the nuances of such processing pipelines. This can result in suboptimal results, frustration, and disillusionment with the value of automated processing. OVAI's tools for ocean scientists will thus need to be easy to operate and maximally transparent. OVAI can provide ML education that will help ocean scientists better understand ML and communicate with their computationally-oriented peers. This can be achieved both with direct teaching materials (e.g., workshops, seminars, bootcamps, etc.) and with human-computer interactions that provide better insight into how a system functions. These efforts will provide domain scientists with the language and tools to constructively use metrics that describe the model performance and data quality, ultimately yielding more constructive collaborations and more effective automated pipelines.  

OVAI can help bring ML expertise into the ocean space by providing evolving, curated benchmark data sets. There is a demonstrable need in the computer science community for dynamic ML benchmarks to enable bleeding-edge experiments and avoid issues with overfitting. These resources could exploit standardized metadata to build repositories that target diverse problems for ML developers. Several systems exist in the ML space that could be used as models for this element of OVAI such as Dynabench, iNaturalist, or WILDS~\cite{van2018inaturalist,kiela2021dynabench, koh2021wilds}. Providing OVAI benchmark resources both as static downloads and with an interactive API will increase interest among computer scientists. 

OVAI will need to support data hosting for users that do not have the means to support public-facing, consistently accessible data servers. This could involve instituting a comprehensive service agreement and implementing a scaled pay model based on processing volume. The rates could be prorated based on open data contributions and annotation verification efforts. Such rates should also be scaled based on the institution in order to support the work of organizations without access to vast financial resources. In the short term, OVAI has begun a partnership with the US National Oceanographic and Atmospheric Administration (NOAA) to provide hosting for annotated visual data, with a 75-year life cycle, for individuals and organizations that cannot maintain consistent public-facing access.  

Ocean science community members pointed to many constraints on their ability to process their visual data, and issues with public-facing storage were a common theme. OVAI should thus be careful to make its tools modular with multiple entry points for users with different needs. Each component must also be scalable and allow for changes in scope depending on the specific project.  

Our one-on-one interview participants came from many corners of the Blue Economy: academic, industrial, government, non-profit, and enthusiasts. However, the interviewees were largely based in the Global North, and mainly in the United States. This is partially a reflection of the authors’ professional networks. It also reflects the state of ocean sciences in general: working at sea is expensive and is typically the purview of wealthy nations~\cite{amon2022my,smith2022deep}. Projects like OVAI seeking to enhance access to ocean data will need to continue efforts to engage these groups and make efforts to ensure their tools are accessible to everyone. The OVAI Portal is intended to be a gateway tool to enable new analysis efforts for many in these underserved regions~\cite{bell2022low}.   

The interviews and workshop revealed a real desire from a dedicated group of ocean enthusiasts to participate in scientific endeavors. This represents an opportunity to harness their efforts for science initiatives while building a knowledge base for ocean researchers to draw on in the future. We are designing a number of entry points for enthusiasts of all levels, from dedicated deep-sea exploration aficionados to grade school students just getting their feet wet. Part of this might entail developing specific, gamified interfaces for these users ~\cite{li2013game,qian2016game}. The interfaces will ideally simultaneously educate and inspire players while generating useful information. Designing such systems will be an HCD effort unto itself, taking care to build interfaces that go beyond an annotation wrapper~\cite{bonetti2021measuring}. Moreover, the OVAI team will need to carefully consider the role of the players to avoid exploitation; an issue in both domain science and HCI circles~\cite{preece2016citizen,hsu2020human}. Critically, any OVAI-related game development will necessarily treat players as a pool of talent worth teaching; many of our enthusiast participants expressed a desire to learn more and directly engage with professional researchers. 

While the insights gleaned in our work will specifically inform the development of OVAI, we believe that our findings and process can be instructive for other projects seeking to assist big science efforts. There is a universe of stakeholders outside of one individual lab and finding solutions to their problems will inevitably make for a stronger, more useful product. In the case of OVAI, this will yield a more accessible tool that streamlines data interactions and enables learning experiences across a spectrum of users.

From the outset, OVAI has sought to build systems that address needs across a spectrum of users united by an interest in the ocean. To make OVAI maximally effective, HCD principles will be critical as the OVAI tools are further developed and distributed; the work presented in this paper is just a starting point. We are implementing a flexible and iterative community co-design process that seeks user input via actual interaction with OVAI tools~\cite{disalvo2009local,hsu2017community}. This will come in the form of surveys, advisory groups, and technical working groups all comprised of people using OVAI. The process will be guided by a disintermediation design rubric to avoid the creation of unnecessary elements and barriers both within and between components of the OVAI suite~\cite{raghavan2017disintermediation}. The rubric will help our team build a system that can promote sustainable practices both from individuals, by connecting their habits to downstream impacts in the ocean, and organizations, by providing easily interpretable data products to inform decision and policy making. Data will be collected about user interaction with OVAI, with the user's permission, to quantify engagement and efficacy to inform design decisions. Our team is currently codifying the rubric and a set of best practices for data collection. These can be used as a general-purpose template for applied HCI projects in scientific data pipelines and for management tools. 

Insights gleaned from our work can be broadly informative to HCI-based development efforts for interdisciplinary science and environmental policy applications. Such design programs are increasingly necessary for science-based decision-making, particularly in the context of the climate crisis, which leverages multiple types of expertise and requires collaboration across traditional academic boundaries. As the volume of data increases, carefully considered systems like OVAI will be critical to effectively use machine learning systems, despite the potential for bias, to address seemingly insurmountable problems involving data collection, annotation, and analysis to inform global policy.

\section{Conclusion}

In this paper, we presented a human-centered design study that details community needs for systems to assess the processing of ocean image data. Our process of interviewing and discussing with participants invested in ocean sustainability and exploration highlights the many needs and use cases for an initiative such as Ocean Vision AI. Collectively these conversations suggest a path forward for OVAI: develop collaborative workspaces that enhance access to ocean expertise, machine learning know-how, and visual data throughout the depth and breadth of the ocean. The infrastructure will need to support and engage each user archetype with: intuitive AI-assisted interfaces; mechanisms to upload, search, and retrieve data based on diverse parameters; resources to facilitate scientific analysis and data storytelling; and visualizations to inform policy decisions. The challenge moving forward lies in ensuring that these tools are easily accessible to all and flexible enough to support different use cases and users.


\subsubsection*{Acknowledgments}
Ocean Vision AI is supported by the National Science Foundation Convergence Accelerator Track E Phase I and II (ITE-2137977 and ITE-2230776). Additional support comes from the Monterey Bay Aquarium Research Institute through generous support from the David and Lucile Packard Foundation. The authors gratefully acknowledge the interviewees and workshop participants. Their insights form the backbone of this paper and had a tangible impact on the design of Ocean Vision AI.

\bibliographystyle{unsrtnat}
\bibliography{ovai-chi.bib}

\section*{Appendix: Interview Template}
Warm-Up ($\sim$5 minutes) 
\begin{enumerate}
\item Can you tell us a little about yourself? Please introduce yourself.
\item What are your motivations for speaking with us today?
 \end{enumerate}
Your Interests \& Goals ($\sim$15 minutes)
\begin{enumerate}
\item What are some of the major scientific / research / conservation / management questions you care about? OR, What is your organization working on in the ocean space? OR, Why are you enthusiastic about the ocean or passionate about ocean science?
\item What methods and techniques do you currently use to answer these questions? OR, What does your organization do? OR, How do you engage with the ocean or ocean science?
\item What hardware and software do you use in your activities? Do you use sensors or imaging technology? If so, which ones?
\item What are your groups’ needs in processing / generating underwater visual data? What observations and/or data products would you like to achieve from your visual data? OR, How would you like to interact and contribute to processing / generating underwater visual data?
\item What are your groups’ limitations in processing / generating / contributing to underwater visual data?
\end{enumerate}
Your Community ($\sim$10 minutes)
\begin{enumerate}
\item Tell me about your collaborator network. Who do you usually work with? (e.g. other researchers, students, community groups)?
\item How do you generally communicate about your work? (e.g. academic papers, Twitter, community conversations, etc.)
\item Where does the funding / support / resources for your work typically come from? [if you are able to share]
\end{enumerate}
Ocean Vision AI ($\sim$15 minutes)
\begin{enumerate}
\item The Interviewer describes Ocean Vision AI.
\item How could an initiative such as Ocean Vision AI address your needs?
\item Do you have any concerns about using AI data in your applications with regards to regulatory requirements, intellectual property, etc?
\item Do you or your organization have any concerns or hurdles to using a framework like Ocean Vision AI for your data (IP, research embargos)?
\item How do you envision your role in an initiative like Ocean Vision AI?
\end{enumerate}
Close ($\sim$5 minutes)
\begin{enumerate}
\item As the interview comes to a close, thank the participant for their time and thoughts and acknowledge how valuable their time and effort is. Make space for the rest of the team to ask any remaining questions, and ask the participant if they have any questions for you, or if there is anything you didn’t cover that they think is important. 
\item Anything else you’d like to share?
\item Is there anyone else we should talk to?
\item Do you have any questions for us?
\end{enumerate}

\end{document}